# The FuturI$_C$T[1] Knowledge Accelerator

## Unleashing the Power of Information for a Sustainable Future

*With our knowledge of the universe, we have sent men to the moon. We know microscopic details of objects around us and within us. And yet we know relatively little about how our society works and how it reacts to changes brought upon it. Humankind is now facing serious crises for which we must develop new ways to tackle the global challenges of humanity in the 21$^{st}$ century. With connectivity between people rapidly increasing, we are now able to exploit information and communication technologies to achieve major breakthroughs that go beyond the step-wise improvements in other areas.*

*It is thus timely to create an ICT[2] Flagship to explore social life on Earth, and everything it relates to, in the same way that we have spent the last century or more understanding our physical world. This proposal sketches out visionary scientific endeavours, forming an ambitious concept that allows us to answer a whole range of challenging questions. Integrating the European engineering, natural, and social science communities, this proposal will release a huge potential.*

The need of a socio-economic knowledge collider was first pointed out in the OECD Global Science Forum on Applications of Complexity Science for Public Policy in Erice from October 5 to 7, 2008. Since then, many scientists have called for a large-scale ICT-based research initiative on techno-social-economic-environmental issues, sometimes phrased as a Manhattan-, Apollo-, or CERN-like project to study the way our living planet works in a social dimension. Due to the connotations, we use the term knowledge accelerator, here. An organizational concept for the establishment of a knowledge accelerator is currently being sketched within the EU Support Action VISIONEER, see www.visioneer.ethz.ch. The EU Flagship initiative is exactly the right instrument to materialize this concept and thereby tackle the global challenges for mankind in the 21$^{st}$ century.

---

[1] Connotations of FuturIcT include "Future ICT", "FIT", and what Wikipedia writes about the usage of the word "futurist": "The term 'futurists' most commonly describes authors, consultants, organizational leaders and others who engage in interdisciplinary and systems thinking to advise private and public organizations on such matters as diverse global trends, plausible scenarios, emerging market opportunities, and risk management."

[2] In this document, ICT stands for Information and Communication Technologies, as usual.

# A: AMBITION

## A1 Goals

The greatest bottleneck of ICT systems today is the difficulty in making sense and efficient use of the large amounts of data we generate. The FuturIcT flagship shall, therefore,

- develop novel ICT systems (incl. applications and infrastructures) combining the best of human and computational abilities to support the understanding, integrative design, and management of complex systems,
- apply these to model techno-social and economic, transport, environmental and other global systems,
- create instruments to support the self-organization, decision-making and governance in politics, business, industry, and academia, with the aim to foster societal goals (e.g. robust techno-social and sustainable economic systems),
- develop principles and tools that will facilitate the emergence of high quality processes, products and institutions in techno-social networks.

The Flagship is oriented at visionary high-risk research, integrating multiple scientific disciplines to a previously unseen extent.

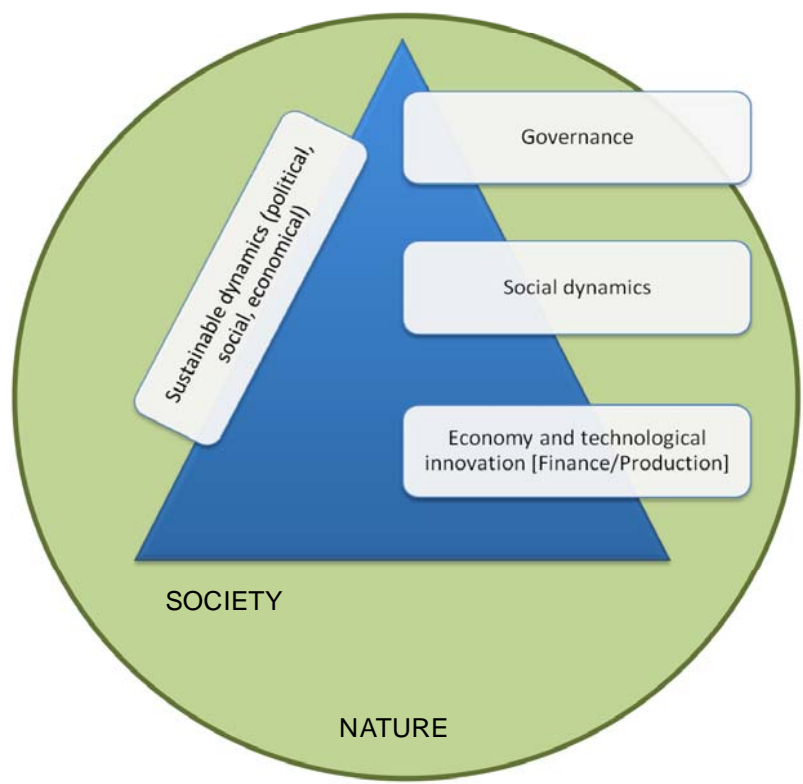

Figure 1. Sketch of the integrative, systemic approach of the FuturIcT Flagship (illustration by Andrea Scharnhorst)

President Lee C. Bollinger of New York's prestigious Columbia University formulated the challenge as follows: *"The forces affecting societies around the world ... are powerful and novel. The spread of global market systems ... are ... reshaping our world ..., raising profound questions. These questions call for the kinds of analyses and understandings that academic institutions are uniquely capable of providing. Too many policy failures are fundamentally failures of knowledge."*

Humankind has reached a situation where existing policy instruments are unable to provide sustainable outcomes on a global scale. It is obvious that our economy and society are facing serious new problems, and it will determine the future of humanity whether we will be able to address them successfully or not. Politicians, business people and scientists need to get into a position to better foresee and proactively moderate the on-going systemic changes rather than having to react to them in a state of crisis. The development of new concepts is urgently needed in order to be able to assess and shape future techno-social and economic systems successfully. To overcome the serious knowledge gaps and weaknesses of our current theoretical understanding of these systems, it will be necessary to make big steps forward.

## A2 Innovation

### A2.1 Grand Socio-Economic Challenges

Ground-breaking ICT research, in intense collaboration with multiple other scientific disciplines, will be crucial to identify the success factors of societies and to address the grand challenges of humanity. This includes
- how to avoid socio-economic crises, systemic instabilities, and other contagious cascade-spreading processes,
- how to design cooperative, efficient, and sustainable socio-technical and economic systems,
- how to cope with the increasing flow of information, and how to prevent dangers from malfunctions or misuse of information systems,
- how to improve social, economic, and political participation,
- how to avoid "pathological" collective behavior (panic, extremism, breakdown of trust, cooperation, solidarity etc.),
- how to avoid conflicts or minimize their destructive effects,
- how to cope with the increasing level of migration, heterogeneity, and complexity in our society,
- how to use environmental and other resources in a sustainable way and distribute them in the best possible way?

Many of the world's challenges cannot be solved by technology alone, but require us to understand the collective social dynamics as roots of these problems and key to their solution.

### A2.2 Need of Massive Data Mining and Reality Mining

- Massive data mining could reduce serious gaps in our knowledge and understanding of techno-social-economic-environmental systems.
- Crises observatories (for financial and economic stability, conflicts, diseases...) could predict crises or identify systemic weaknesses, and help to mitigate impacts of crises.
- Real-time sensing and data collection ("reality mining" of flu infection data, gross domestic product, environmental data, cooperativeness, compliance, trust, ...) could reduce mistakes and delays in decision-making, which often cause inaccurate and unstable control

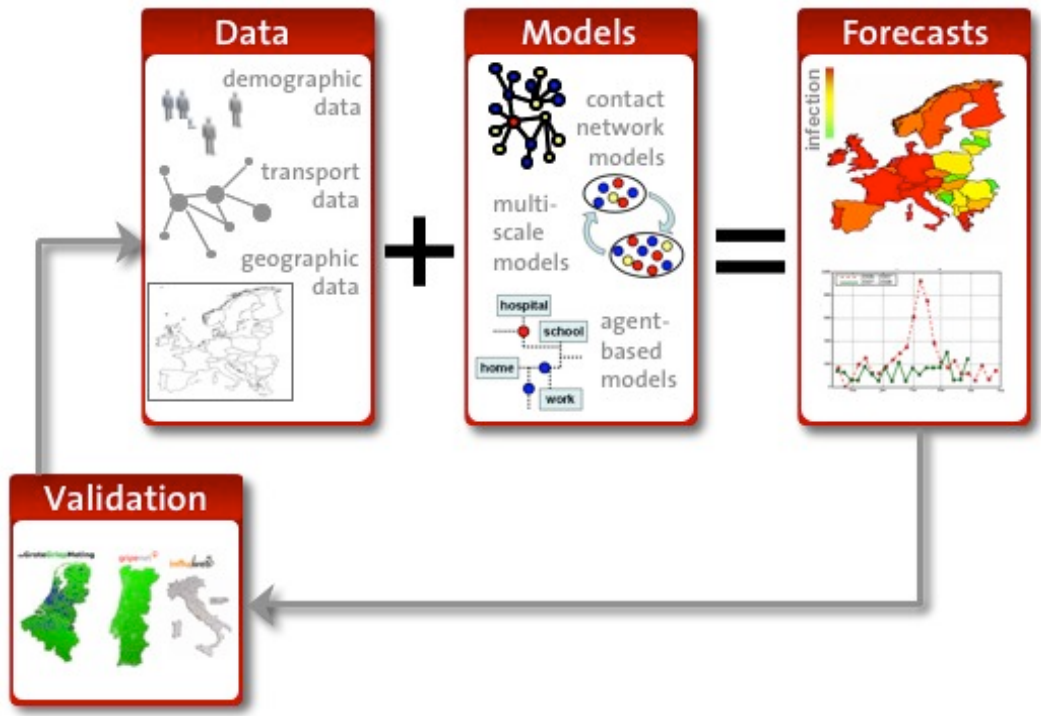

Figure 2. Illustration of the systemic modelling approach within the FuturIcT Flagship (illustration by Alex Vespignani)

**A2.3 Need of Social Super-Computing**

Besides massive data mining capabilities, it is required to build up suitable super-computing capacities for the simulation, optimization, and management of sustainable techno-social and economic systems. Gigantic computer power is, for example, needed for large-scale computational analyses in the following areas:

- Massive data mining, e.g. real-time financial data analysis
- Network research, community detection
- Monte-Carlo simulations of probabilistic system behavior
- Multi-agent simulations of large systems (e.g. "whole earth simulation", which may involve up to 10 billion agents and complementary environmental simulations)
- Multi-agent simulations considering human cognitive and psychological processes (e.g. personality, memory, strategic decision-making, emotions, creativity etc.)
- Realistic computer simulations with parameter-rich models (coupling simulations of climate and environmental change with simulations of large techno-social-economic-environmental systems)
- "Possibilistic" multiple world-view modelling (to determine the degree of reliability of model assumptions and to improve the overall prediction capability)
- Calibration of parameter-rich models with massive datasets
- Scanning of multi-dimensional parameter spaces
- Sensitivity analyses (e.g. $k$-failures)
- Parallel worlds scenario analyses (to test alternative policies etc.)
- Visualization of multi-dimensional data and models of complex systems
- Optimal real-time management of complex systems ("guided self-organization", "self-optimization")

It should be underlined that most challenges addressed by the FuturIcT Flagship concern a *combination* of all the above points!

**A2.4 Expected Outcomes of the FuturIcT Flagship**

The FuturIcT Flagship will build a **knowledge accelerator to address the challenges for humankind in the 21$^{st}$ century.** This will include the creation of the following:

- Integrated data collection, modelling, simulation, visualization, design, and decision-support tools for complex systems, based on novel approaches such as "possibilistic" simulation concepts (multiple world view, parallel world scenarios,…)
- Social super-computing, including a socio-economic modelling language
- An innovation accelerator changing the scientific production paradigm, to support efficient progress and investments in science and technology
- A peer-to-peer reputation and privacy-respecting recommender system
- "Knowledge engines" combining real-time data collectors with social information theory to harvest the knowledge of humanity
- Crisis observatories (for financial and socio-economic instabilities, conflicts, epidemics, environmental changes, etc.)
- Individually customizable reputation and recommender systems that respect privacy
- A novel theory of socio-economic robustness and an understanding of "social pathologies" (riots, panic, etc.)
- A **Living Earth Simulator** for global-scale simulations involving interactions of up to 10 Billion agents, coupled to a simulated and/or measured environment
- Situation-room-like policy simulation and visualization centers ("decision theaters" for policy decisions)
- The tools required for an **international socio-economic crisis management center,** including a whole earth and policy simulator, plus a contingency and resilience toolset, available for everyone.
- Showcases (demonstrators) regarding the global financial system, sustainable transport, energy generation and production, epidemic forecasting, and a disaster and evacuation management system, ...

**A2.5 Paradigm Shifts Expected from the FuturIcT Flagship**

- From pre-defined to self-organized ICT systems
- From searching to discovering
- From data fitting to reverse engineering
- From ubiquitous computing to social computing
- From a web of data to a web of models
- From exclusive to possibilistic models
- From particular to systemic models (a global view from the start)
- From information processing to knowledge production**,** from knowing to understanding
- From solving to anticipating problems
- From producers and consumers to prosumers (participatory consumption, co-creation, individualized products), from supply chains to demand chains
- From quantity to quality
- From an economic to a socio-ecological approach (considering that social exchange is multi-faceted and not only driven by financial forces; oriented at a robust, sustainable financial and economic system with social and economic inclusion of everyone)
- From improvised crisis management to real-time decision systems
- From information communication technology to imagining, computing, transforming

## A3 Comments

**A Quote from Josh Epstein (Brookings Institution, USA):**
*"We are poised at the cusp of interacting epochal changes: ICT is propelling humanity into the age of global human connectivity; we are changing the global environment; we are peering into the human genome and unraveling the neurochemistry of human emotion and behavior. ICT is at once propelling these changes, but also permitting us to comprehend them. Planetary-scale computational modelling is now feasible, allowing the study of coupled transitions at multiple scales.*

*These epochal changes eclipse the turbulence of daily political affairs. And their complexity dwarfs the capacity of any individual's comprehension. Only a collective mind enabled by the ICT resources of our [the Flagship] consortium can undertake credible actionable forecasts embracing all of this, for the first time, in a rigorous replicable manner. And it is imperative that this admittedly bold step be taken: to envision—as comprehensively as the best minds and best ICT permit—how these epochal developments will interact over the next decade. The coupled socio-economic-environmental dynamics will [be] far from linear, far from equilibrium, and far from canonically rational. But they can be understood, and productively shaped, by the Flagship proposed here. It is an experiment we can't afford not to do."*

The FuturIcT Flagship will simulate the complex emergent system that society is, and the relevant systems it is embedded in.

Society emerges from the myriad of interaction among its members and with its environment. People are heterogeneous, goal driven consumers and producers of both, resources and information. We endeavor to reverse engineer modern society and, in doing so, come to better understand the emergent processes that have (and will have) profound impacts on people's lives.

The following paragraphs outline the fundamental contributions of the FuturIcT Flagship towards this goal, while the economically and practically relevant challenges are addressed in the Section "Impact".

## A4 Methodological ICT Challenges

There are a range of methodological challenges that need to be faced. We list here some of the main issues that have been highlighted

**Exascale Computing and Living Earth Simulator:** Empower exascale computing through fundamentally new algorithms, data structures and compilers capable of exploiting the massive parallelism of future computational systems; create a global-scale multi-level Living Earth Simulator with large realism (10 petaflops in 2013, 100 petaflops in 2016, 1 exaflop in 2019).

**Highly Decentralized and Peer-to-Peer Systems:** Develop powerful algorithms to make peer-to-peer (P2P) systems trustable and resource-efficient; explore new kinds of P2P applications (e.g. micro-finance, flexible self-control, etc.); create collaborative ICT systems that can satisfy large-scale computational needs without requiring centralized facilities and institutions.

**Reality Mining:** Develop concepts for large-scale long-term real-time data gathering and sensing "on the fly"; measure systems and behavior in real time using mobile phones, GPS, accelerometers, RFID tags, search engine requests, web and email use; create sensor networks/smart grids; employ opportunistic sensing; learn how to collect data from users in ways that protects privacy; develop suitable methods to generate surrogate data having the same statistical properties as the original data, but do not allow the identification of individual people.

**Swarm Computing:** Develop highly distributed, adaptive processing and storage based on principles of collective intelligence; employ emergent computing and artificial immune systems; make swarm architectures programmable like single systems; create ICT platforms that can reconfigure themselves and adaptive algorithms that re-program themselves, to reflect/allow for changing interaction rules; develop a standardized framework for modelling agent interactions; consider cognitive and emotional factors supporting swarm intelligence; enable "cultural evolution" in computer networks.

**Social Computing:** Conceptualize and analyze social computing as the interaction between digital and social networks; improve integration of human wisdom into ICT systems ("human-computer confluence"); automate ratings of contents by "emotional sensors"; empower ICT systems to sense disagreement in order to learn what the user wants it to do or not to do; develop tools to improve human-computer and ICT-mediated interactions considering human affects and emotions; identify dangers and utilize potentials of collective decision-making and intelligence ("wisdom/madness of crowds", "prediction markets", etc.); allow users to coordinate efforts in on-line communities and steer the collective activity towards predefined goals ("hive mind").

**Social Information Theory:** Learn how to distinguish meaningful, relevant, or transformative information from unimportant information; understand the role of social, affective interactions for the meaning and impact of information; identify relevant variables, parameters, and stylized facts of complex systems; extract knowledge from information by suitable adaptive filtering techniques.

**User-Oriented ICT Systems:** Create "non-expert systems", which make the knowledge of humanity, simulation tools etc. accessible for everyone; design models for adaptive autonomous systems that interact with humans and social structures: ICT systems that think like humans, understand what they want, and adapt to them; develop ICT systems that are intuitive, easy to use, customizable, reliable (self-repairing), flexible (self-adaptive), trustable (in particular self-protecting), sustainable, scalable, and interoperable (like plug in's or App's); facilitate an "interface-free" interaction with users.

## A5 Applied ICT Challenges

**Data Collectors:** Support data collection, fusion, filtering, and categorization; suppress pollution of data bases and information systems; improve accessibility and data extraction; support massive techno-social-economic-environmental data mining; manage data deluge; improve handling of inconsistent data; perform model calibration, identifying and filling data gaps (e.g. in case of incomplete or non-representative data); use electronic and sensor networks for automated data collection and hypothesis testing; measure regional differences in social behavior; establish "moral sensors" and "compliance detectors" to identify changing norms; enable crowd sourcing and the participation of the population in political decision-making ("eGovernance").

**ICT-Empowered Systems Modelling:** Support non-theory-based, data-driven discovery; develop methods of model-enhanced data representation and interpretation; facilitate fast model prototypes; perform sensitivity analyses to model assumptions, structural and parameter variations; identify early warning signs of upcoming critical states and systemic shifts; extract a variety of mathematical laws (including multi-level dependencies) that are consistent with available data sets; evaluate their validity, simplicity, and sensitivity; assess the suitability to interpret these models, to perform analytical studies, and to calibrate them; simplify models to determine stylized facts of the studied system; identify well measurable, interpretable and relevant variables; customize the degree of detail to the modelling purpose; determine implications that distinguish alternative models and derive procedures to test model variants.

**Evaluating ICT Systems:** Evaluate and rank models according to individual criteria; implement cognitive processes in ICT systems; support multi-dimensional evaluation of reputation-relevant aspects; facilitate customized and community-specific reputation; allow coordination and support cooperation between individuals and stakeholders; simplify community detection and the formation of groups with compatible quality standards/values.

**Reasoning ICT Systems:** Complement, enhance, and fuse data mining algorithms with models and expert knowledge to create interpretation devices; develop methods for the self-validation of algorithms and models; run potentially relevant scenarios; identify causality chains; explore feedback and cascading effects for all model variants; determine the reliability of implications, given the validity of the underlying models; compute, analyze and use independent information to address the impact of model assumptions and unmask hidden constraints.

**Creative ICT Systems:** Support the identification of open points and crucial questions regarding data analyses, algorithms, and models; develop a hypothesis generator; use evolutionary computing and develop imaginative ICT methods to test alternative interaction mechanisms and develop integrative systems designs; identify governance options in techno-social-economic-environmental systems; support the creation of propositions, brain storming, and decision making.

# B: IMPACT

The FuturIcT Flagship will create interactive, multi-purpose modelling, exploration, and systems design tools that use the best combination of human and machine intelligence. Experts will be able to choose among the variables, parameters, model variants, simulation scenarios, hypotheses to be explored, and system designs proposed by this semi-automated tool (the "knowledge accelerator"). It will stimulate creativity and extend the limits of imagination. The knowledge accelerator will also provide a living, self-organizing data pool and a cyber-infrastructure for people with various backgrounds.

The integrated, large scale modelling of complex techno-social-economic-environmental systems will promote cross-pollination between different fields. Simplifying and standardizing data collection, modelling and simulation will stimulate interdisciplinary research questions, collaborations and discoveries. Best practices will emerge and shared standards of evidence will result in fundamental and substantial advances.

Using crowd-sourcing and reputation systems, an innovation accelerator will be built to support the collaboration of the best fitting partners. It will help to separate relevant from irrelevant information, thereby increasing the transparency, usefulness and control of the information everybody is exposed to, and it shall be able to deal with inconsistencies by considering multiple model variants and to determine their respective validity. The FuturIcT Flagship will simplify scientific research and evaluation procedures, generating higher-quality outcomes. The application to social and economic systems will deliver a new economic theory that captures realistic human behavior and system states far from equilibrium.

The Flagship is expected to cross-fertilize developments in multiple areas and to improve the capacity to cooperate. Through a free sharing and better accessibility of massive data sets and multidisciplinary scientific knowledge, it will kick off a new age of systemic modelling and simulation, triggering radical innovation in all areas of society, technology and economics. It will provide the tools to make humanity fit to cope with the problems of environmental and demographic change, health, safety, security, and sustainable development. Furthermore, taking into account affective components of information transfer will improve the quality of human interaction with engineered systems, including ICT systems.

The innovations of the FuturIcT Flagship will be practically relevant for many sectors of society. In fact, information has become a critical resource in the economy. The FuturIcT flagship will support the communication with customers and end users, and the development or even co-creation of highly customized products. This will allow a wider variety of interests and agents to be expressed and satisfied. In this way, the FuturIcT Flagship will create new business opportunities, particularly for small and medium-sized enterprises. However, FuturIcT will also help to reduce losses by financial and other crises, inefficiencies in transportation and production systems, and social problems, thereby saving large amounts of tax payers' money.

## B1 Future Living

The fast and wide access to information has changed society on all levels: individuals, groups, companies, and governments. This circumstance is reflected by terms such as "information society" and "knowledge economy". Never before has it been so crucial to transform information into knowledge efficiently.

**Customized Information Services:** Support information discovery, evaluation, and integration; evaluate information, projects, cooperation partners, or products, based on objective quality criteria and subjective ratings; offer individual decision-support by providing multi-criteria, customized, privacy-respecting recommendations; enable quality emergence and diversity-oriented navigation through multi-dimensional quality landscapes, highlighting alternatives; detect changes in user/customer interests and tastes in a timely manner; support individualized services of all kinds; provide the right information at the right time and the right place as efficiently as possible; suppress unwanted, unneeded, or "polluted" information; archive information effectively for future use; pave the way for a quality-oriented reputation society.

**Innovation Accelerator:** Support a quick recording and dissemination of ideas, methods, and data; implement new publication concepts ("Science2.0"); develop more efficient ways of scientific co-creation and cooperative quality/value production; support collaboration through new ICT tools; analyze scientific activities; reveal new trends and collaboration networks; identify leading scientists and institutions through novel, multi-dimensional performance measures; study career paths (brain drain); simplify evaluation of innovation, quality, and performance; disseminate knowledge beyond circles of specialists.

**Personalized Education:** Provide ICT platforms (including "serious game" environments) to enhance individual and collective learning; support efficient sharing of information, scenarios, and best practices; collect depersonalized database of individual "learning trajectories"; apply collaborative filtering methods and extend recommender systems for educational means; enhance learning by linking with institutions (libraries, archives, etc.), experts, lay-specialist knowledge, and folksonomies.

**Smart Cities, Transport, Traffic, Logistics:** Measure real-time travel activities and their environmental impacts in a privacy-respecting way; provide better planning tools for coordinated and environmental-friendly travel activities and logistics; develop and implement more flexible, efficient and scalable control approaches for transport and logistic systems; improve concepts and user-friendliness of multi-modal transport; understand the interconnection between traffic and land use and consider them in urban and regional planning.

For the first time in our history, the majority of humanity is living in an urban setting and this proportion will increase in the future. Congestion and property price bubbles are symptoms of coordination problems. Congestion generates losses of productive time, wastes energy, and pollutes the environment. It creates economic losses of 10 Billion US$ each year in the US alone. Increasing the capacity of the transport system by 5 to 10% by better coordination could possibly reduce cumulative delay times by up to 50%, amounting to several Billion US$ in the US alone. The related reduction of $CO_2$ emissions would be significant as well.

**Smart Energy Production and Consumption:** Elaborate new coordination schemes for highly decentralized energy production and consumption in "smart grids" with a large number of generators and loads; optimize the generation, delivery, and electrical grid structure; identify the behavioral laws of electricity producers and consumers (periodicities and extremes); develop new incentive structures to match supply and demand; work out concepts to promote local and diverse energy production and consumption; elaborate ways for a smooth transition from nuclear and coal-based energy production to more environmental-friendly and sustainable ones; study ways, in which people can be stimulated to reduce energy consumption and invest more into sustainable technologies to meet $CO_2$ emission goals.

**Safety and Security:** Find efficient solutions to reduce corruption, crime, conflict, and war (including civil war); understand asymmetric multi-party conflicts and spillover effects; predict emerging conflicts from complex interdependencies; find solutions to integration problems resulting from migration; explore new responses to the problem of independence/liberation movements and the conflict in the Middle East; test and apply theories and technologies to monitor the spreading of crime and corruption within and between societies; fight criminal (e.g. terrorist, drug dealing and fraud) networks by studying the growth characteristics and structure of known criminal networks; develop techniques to reconstruct unknown parts of the network; identify the most efficient and effective ways to dismantle these networks; apply strategies to disrupt internet crime (including malicious bot networks).

## B2 Towards Robust and Sustainable Systems

One of the goals of the following research streams is to derive general principles of how short-term needs can best be met while balancing them with long-term constraints.

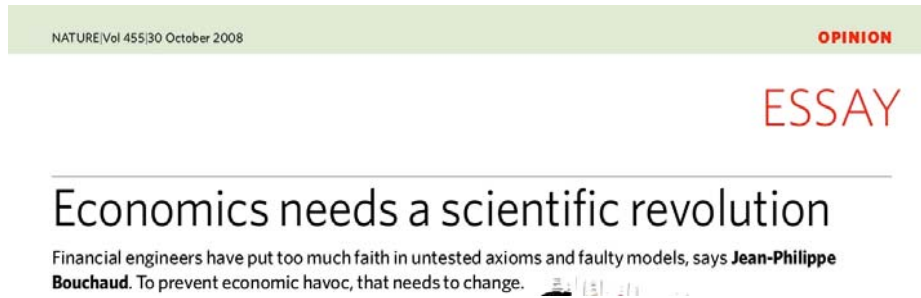

**Realistic Theory of Economic Systems:** Develop a realistic economic theory to give more reliable advise to decision-makers; go beyond the paradigms of the "Homo Economicus" (the "perfect egoist"), efficient markets, equilibrium models, and representative agent models (mean-field models); develop agent-based models of boundedly rational behavior (e.g. limited cognitive capacities, behavioral biases and emotional aspects); consider randomness, extreme events, heterogeneity, non-linearity, emergence, and complexity to yield a greater descriptive and predictive model validity; make models consistent with empirical and experimental data; work out the mathematical connection between microscopic and macroscopic economic theories.

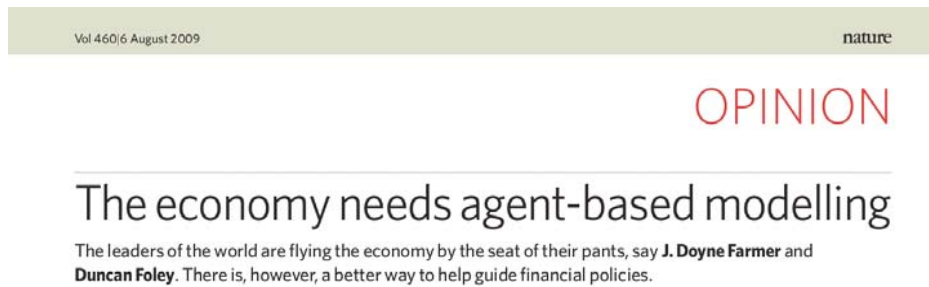

**Crisis Observatories:** Implement massive data mining and suitable filtering techniques to detect forth-coming or possible crises, e.g. bubbles or crashes in financial or housing markets ("market monitoring"); identify advance warning signs for financial and economic instabilities, for shortages in supply (oil, water, food, …), wars and social unrests, epidemics, environmental change, etc.); extract laws of systemic instabilities; identify interdependencies, feedback loops, and causality chains that may lead to cascade spreading effects.

**Contingency Plans and Risk Management:** Perform crises prediction and containment, detecting feedback loops and possible cascading failure before they happen and cause wide-spread damage; model individual and social behavior under conditions of disasters/crisis; develop new ICT concepts for an adaptive risk management; facilitate and support better disaster preparedness and response management; create tools for the simulation and a prompt planning of large-scale evacuation operations; develop crises management concepts considering the behavior of complex systems under uncertainty.

**Managing Complexity and Institutional Design:** Support the coordination between multiple parties; study systemic effects of over- or under-regulation; identify the optimal mix of central and decentral control; find methods to support a consensus or balance between different interest groups and institutions; explore new ways to overcome political blockages in situations with incompatible interests; learn how to utilize self-organization principles to reach system-optimal solutions ("guided self-organization"); find management strategies that support a robust, but flexible and adaptive organization; identify novel conflict resolution mechanisms; develop new ways to increase social and economic participation (e.g. "eGovernance").

**Integrative Design of a Sustainable Financial System and Economics:** Explore mechanisms and institutional settings, which can create robust, but adaptive social, political, and economic systems under conditions of global change; develop a non-equilibrium theory of economic exchange with and without central markets and reference currencies; model value as emergent property, considering psychology and social interactions; compare properties (like efficiency, predictability, reliability, robustness, signs of failure, etc.) of alternative banking, pricing, auctioning and market systems by extensive computer simulations; investigate different value transfer systems; study distributed and open kinds of credit and exchange systems (e.g. P2P lending, P2P banking); facilitate a more efficient and robust exchange of value between agents, considering mechanisms like trust, reputation, and norms; support peer-to-peer economic exchange between producer/seller and customer; explore financial policies and new institutional or regulatory settings by computer simulations in order to support decision-making.

On the one hand, this research stream is expected to be crucial to avoid future economic crises or at least mitigate them. On the other hand, the economic potential in terms of value creation could eventually amount to many billion EUR.

**Global System Dynamics and Policy:** Elaborate ICT-based methods leading from data, through models to policy and decision making; create techniques, tools, and concepts allowing the integration of system component models into global-scale systemic models of the living earth; develop new possibilities to gather and incorporate data into models and allow others to work with these data; produce ways of verifying the results of large-scale, integrated, systemic models to capture emergent phenomena correctly and avoid artifacts; uncover counter-intuitive interdependencies by computer simulations; improve the visualization of multi-dimensional dependencies in global models to strengthen intuition and decision-making; provide guidelines for model-based decision support; find ways to consider decision-making constraints in model scenarios; support communication and exchange between scientists and decision-makers; study impacts (such as migration, economic transformation, urban development, social unrest, or conflict) of environmental changes, natural hazards, economic development, exploitation of resources, demographic changes, and social interactions (collective behavior); investigate options to address the problems and satisfy the interests of the developing countries more successfully, considering the local and world-wide social and environmental impacts.

# C: INTEGRATION

To a previously unseen extent, the FuturIcT Flagship will promote the integration between

- different scientific disciplines,
- research institutions in many countries of the EU and worldwide,
- research and education,
- academia, business, governance, and the public media,
- scientists and non-expert users, and
- people of different social, economic, and cultural backgrounds, gender, religion, and age.

**C1 Communities and Concepts**

The FuturIcT Flagship will bridge between the natural, computational, engineering, and social sciences, humanities, cultural studies, political science, economics, finance, and many more disciplines.

Note that the following list is not meant to be exclusive or complete. Moreover, many of these concepts are used across disciplinary boundaries, but mentioned here just once.

- Computer science (supercomputing, grid computing, distributed systems, human-computer interaction, semantic web, data mining, sensor networks, ambient intelligence, information systems, database design and management, algorithm design, machine learning, automated deduction, visualization, serious games)
- Mathematics (statistics, modelling, catastrophe theory, non-linear dynamics, extreme events, sensitivity analysis, logic, axiomatic deduction)
- Physics (complex systems theory, self-organization, theory of critical phenomena, power laws, chaos theory, network analysis, experimental data, information theory)
- Engineering (agent-based modelling, scenario modelling, cybernetics)
- Cognitive science, psychology (subjectivity, individual preferences, attention, creativity)
- Sociology (social interaction, cooperation, reputation, norms)
- Political Sciences and Law (governance)
- Anthropology (cultural studies)
- Economics (financial markets, systems design)
- Biology (neuroscience, perception, computational biology, epidemiology, evolution)
- Geosciences (global warming, prediction of natural hazards)
- Ecology (environmental data, sustainability)

The integration of these many disciplines is facilitated by several scientific organizations of interdisciplinary working scientists and supported by COST and Support Actions as well as ERANETS of the EU (see Section "Plausibility").

**C2 Infrastructures**

The FuturIcT Flagship will involve leading research institutions from all over the EU (e.g. Austria, France, Germany, Great Britain, Hungary, Italy, Netherlands, Norway, Poland, Portugal, Spain, Sweden), from partner countries (Israel, Switzerland) and from all over the world (e.g. Argentina, China, Japan, Mexico, USA, Chile).

This includes international **top institutions** such as Cambridge University, University College London-UCL, Oxford University, ETH Zurich, Imperial College, Centre National de la Recherche Scientifique, EPFL, Max Planck Institute for Mathematics in the Sciences, Max Planck Institute for the Physics of Complex Systems, La Sapienza University of Rome, University and TU Munich, Potsdam Institute of Climate Research, Central European University, Collegium Budapest—Institute for Advanced Study,

Institute des Systemes Complexes, ISI Foundation—Institute for Scientific Interchange and Politecnico di Torino, King's College London, London Business School, Stockholm University, Universitat de Barcelona, University of Amsterdam, Warsaw University of Technology, the Open University and Warwick University, to mention only a few. Many of these institutions have super-computing centers, some of them among the fastest ones in Europe (e.g. ETH Zurich).

In the **USA,** researchers from various institutions have indicated their strong interest to collaborate with FuturIcT as well. They are located at the Brookings Institution, Indiana University, Northwestern University, Santa Fe Institute, Stanford University, and the University of California at San Diego.

Over 200 scientists from many more institutions, several of them known to be **world experts** in their fields, have already indicated their interest to participate in the FuturIcT Flagship.

There is also a large interest of business partners, foundations, government agencies, and the media in an involvement into the FuturIcT Flagship. Details of this interaction are currently worked out. Most remarkable at this moment is the support by **George Soros** and the **Institute of New Economic Thinking** that he established with an endowment of 50 Mio. US Dollars. Furthermore, the FuturIct Flagship is negotiating with Gap Minder, PLoS, the Wikipedia Foundation, and other relevant organizations about a privileged collaboration.

**GEORGE SOROS**

March 30, 2010

To Whom It May Concern,

On behalf of the Institute for New Economic Thinking and Central European University I am writing to express strong interest in this scientific endeavor and in collaborating with the candidate flagship FuturICT and the team Professor Helbing is creating.

The Institute for New Economic Thinking (INET) www.ineteconomics.org has been founded to foster and create new interdisciplinary ways to address social and economic problems. Applications of network theories to system evolution, political-economic interactions and psychologically sophisticated approaches to understanding system dynamics are just a few dimensions of exciting new research that our fellows will be working to develop.

Central European University, INET and a number of leading universities are working to establish a network of campus based joint venture institutes around the world to further invigorate our research agenda. The first of which, in conjunction with the Oxford Universities 21st Century School will begin to operate shortly. This interdisciplinary network will add further strength and depth of scholarship to the pursuit of new and deeper understanding of a myriad of social issues.

The team of scientists that Dr. Helbing has gathered together can, I believe, make a significant contribution to the understanding of the evolution and change in societies as they meet the formidable issues of governance, climate change, sustainable economic balance that we are all faced with in the coming decades. I look forward to CEU and INET joining with FuturICT to address these daunting challenges in the coming years.

Sincerely yours,

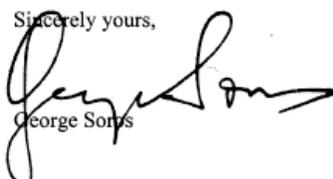

George Soros

# D: PLAUSIBILITY

**D1 Preparatory Steps**

To prepare for the FuturIcT Flagship, the following preparatory steps have been taken:

- A well-connected multi-disciplinary community has been built up.
- Recently, leading journals like Nature and Science have reflected many times upon the need of a bold research initiative addressing socio-economic challenges. The scientific community is prepared and publicly listened to.
- The grand scientific challenges have been identified.
- The complex systems and social simulation community (CSS, ESSA, ASSYST) has created links to the global system dynamics and sustainability community (GSDP project).
- Suitable institutional settings for a large-scale, goal-driven research initiative are currently elaborated (EU Support Action Visioneer).
- The need for social data-mining and crises forecasting capacities is figured out.
- The concept for an innovation accelerator is developed.
- The need for social simulation capacities and integrative systems design centers is worked out.

**D2 What Has Changed?**

The FuturIcT Flagship is not the first attempt to address the challenges of socio-economic systems on a global scale. The most well-known example is the system dynamics approach by the Global 2000 study and the Club of Rome, which have raised the awareness for global issues, but have been limited in their success regarding the quality of prediction.

However, in the recent two decades, a variety of new methods and tools have been developed, which allow one to go considerably beyond previous modelling attempts. The success of the FuturIcT endeavor will be based on a *combination* of the following advances:

- ICT systems are becoming powerful enough to perform global scale simulations.
- There much more test cases to validate models, and further data required for substantial simulations are becoming available through multi-source massive social data mining (e.g. reality mining with mobile phones, sensors, and on the Web).
- Machine learning facilitates to determine significant patterns ("stylized facts") in the data.
- Geographic information systems support the visualization and interpretation of spatio-temporal patterns of global change.
- Environmental science is providing a much more accurate understanding of global and regional scale processes over a large range of time scales and is giving increasingly accurate projections of environmental change.
- Statistical physics, in particular the theory of critical phenomena, phase transitions, and extreme events make it possible to theoretically understand surprising behaviors of complex systems.
- Non-equilibrium statistical mechanics is providing new conceptual tools for studying the response of non-equilibrium system to perturbations.
- The theory of complex networks and networks of networks allows one to grasp complex systems, particularly the coupling of structure and dynamics, of function and form etc.
- A "Hilbert program" of grand fundamental challenges in socio-economic research has been worked out.
- A quickly growing community of interdisciplinarily working researchers from the social, natural, engineering, and computer sciences is available to perform large-scale integrative systemic analyses, combining the best of all knowledge.
- In view of global challenges like the financial crisis, the need for systemic, multi-disciplinary

approaches is now widely recognized.

SOCIAL SCIENCE

## Computational Social Science

David Lazer,[1] Alex Pentland,[2] Lada Adamic,[3] Sinan Aral,[2,4] Albert-László Barabási,[5] Devon Brewer,[6] Nicholas Christakis,[1] Noshir Contractor,[7] James Fowler,[8] Myron Gutmann,[3] Tony Jebara,[9] Gary King,[1] Michael Macy,[10] Deb Roy,[2] Marshall Van Alstyne[2,11]

A field is emerging that leverages the capacity to collect and analyze data at a scale that may reveal patterns of individual and group behaviors.

PERSPECTIVE

## Economic Networks: The New Challenges

Frank Schweitzer,[1*] Giorgio Fagiolo,[2] Didier Sornette,[1,3] Fernando Vega-Redondo,[4,5] Alessandro Vespignani,[6,7] Douglas R. White[8]

The current economic crisis illustrates a critical need for new and fundamental understanding of the structure and dynamics of economic networks. Economic systems are increasingly built on interdependencies, implemented through trans-national credit and investment networks, trade relations, or supply chains that have proven difficult to predict and control. We need, therefore, an approach that stresses the systemic complexity of economic networks and that can be used to revise and extend established paradigms in economic theory. This will facilitate the design of policies that reduce conflicts between individual interests and global efficiency, as well as reduce the risk of global failure by making economic networks more robust.

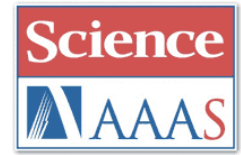

**D3 Scientific Communities FuturIcT Can Build on** (see letters of support)

The FuturIcT flagship brings research communities from the engineering, natural, and social sciences together, to combine and integrate the best of their knowledge:

- ICT-focused PANORAMA (pervasive adaptation) research agenda group (PerAda, ca. 650 members)
- Complex Systems Society (CSS, ca. 2.000 members) and Reseau National des Systemes Complexes (RNSC), collaboration with the Santa Fe Institute
- Institute of New Economic Thinking established by George Soros
- European Social Simulation Association (ESSA, ca. 370 members)
- Physics of Socio-Economic Systems Division of the DPG (ca. 350 members)
- COST MP0801: Physics of Competition and Conflicts (ca. 300 members)
- COST Transport and Urban Development (TUD)
- Sustainability-oriented Global System Dynamics and Policy communities (GSD, GSDP, >200 members)
- Socially Intelligent ICT (ASSYST and COSI-ICT programme, > 100 members)
- European Conference of Transport Research Institutes (ECTRI)
- Open University (with an educational program in complexity science)

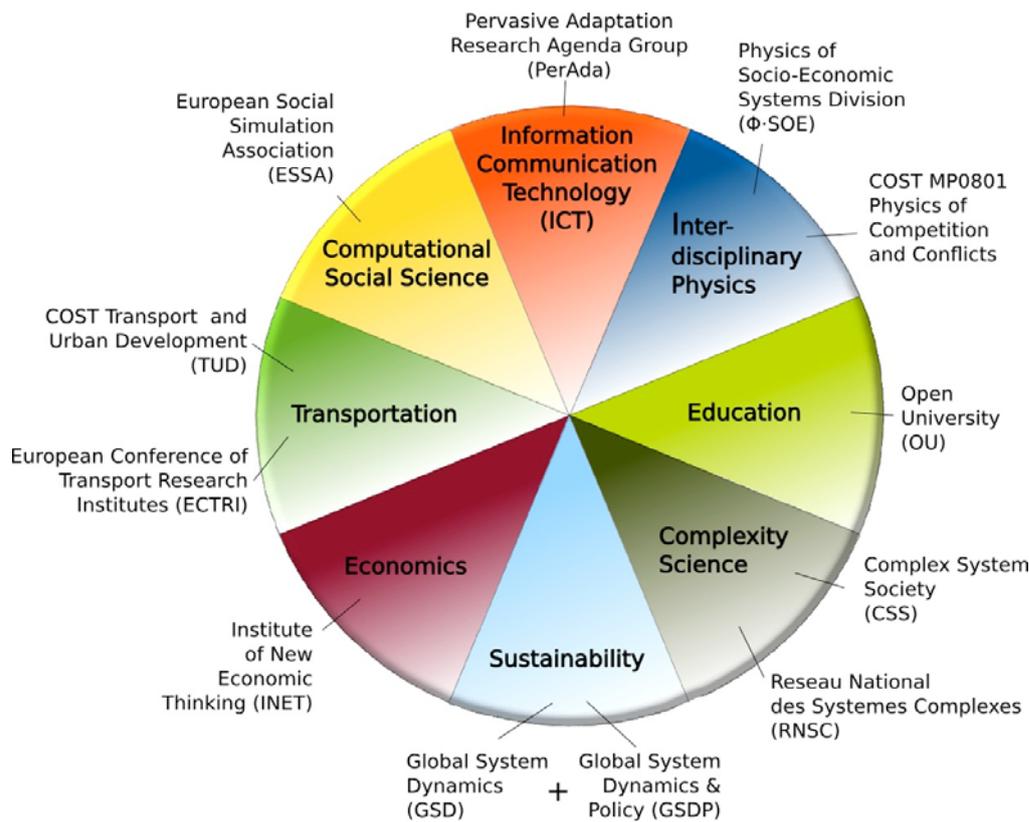

Figure 3. Schematic diagram of interacting research areas (illustration by Stefano Balietti)

**D4 Leverage Effect**

The challenge to leverage 80% of funds from other sources will be a major challenge. However, investing 1 billion EUR into this Flagship is little compared the losses that the last financial crises has caused (about 14.000 billion US Dollar), and also little compared to the investment into recent elementary particle experiments at CERN (9.4 billion EUR) or the Apollo project (about 23 billion in 1969 US Dollars). As explained in the section "Impact", the FuturIcT flagship will also create economic value by many new business opportunities.

As a next step, we will contact the national funding agencies with this Flagship proposal and the letters of support we have, to lobby for supplementary financial support of this large-scale European research initiative on a national level. Related EraNets will help with this process. At the same time, we will contact business partners and government representatives. The support of George Soros will be very helpful in this process, as well as the expected media reports. For this, we have created a contact list of journalists, who have reported about our work in the past.

Furthermore, we are currently discussing the strategy of introducing different categories of memberships. For example, on the side of the researchers, we may require them to show that they have obtained substantial research grants on related subjects, in order to get particular privileges (e.g. the use of special functionalities of the innovation accelerator). In the open funding schemes, this would largely reinforce the research trend that we are creating with the FuturIcT Flagship. However, it is anyway expected that a goal-oriented research initiative like this, involving hundreds of scientists regularly meeting at international workshops, will create a "herding effect" (in terms of attracting more and more people to work this subject). It is known that people start working on new problems if this allows them to increase their scientific impact. We know that many people who have been studying inanimate matter in the past are inclined to tackle the challenges of humanity, if only the institutional obstacles in this research area are removed. This dynamics in itself is expected to create a large leverage effect.

On the side of business partners and foundations, we may have differentiated memberships as well, depending on how much money they spend on related research. For example, category A members could be required to spend a certain percentage of their business volume, and category AAA members would be expected to give considerable donations in exchange for publicity.

It can be noted that we expect FuturIcT to attract a lot of support from business and government agencies. Most large companies such as supermarkets, retailers and telecoms suppliers have huge databases of customer information and routinely use data mining to run their businesses. For any one of these, access to the enormous cutting-edge expertise of FuturIcT would be of great advantage to their R&D departments, certainly exceeding an investment of a few hundred thousand Euros per year as AAA members. The same applies to companies, which make their business with the internet, the world wide web, and sensor technologies. Similarly, banks, insurance companies and other financial institutions could hardly ignore this unprecedented source of scientific expertise that could help them avoid the problems of the recent past and the unknown future. The FuturIcT Flagship initiative will also be highly relevant for urban and transportation planning, and the energy, health, and security sectors. It is, therefore, confident that it can raise financial support for this highly applicable Flagship research program amounting to a multiple of the contribution by the European Commission.

**D5 Examples of Related, Preparatory Research Projects and Fields**

- ERANETS: ComplexityNET, Chistera (under discussion)
- Coordination Actions: Exystence, GIACS, ONCE-CS, ASSYST, PANORAMA/ PerAda, AWARE (pending), GSDP (pending)
- FET and NEST projects: EURACE, EMIL, PERPLEXUS, PATRES, MMCOMNET, EVERGROW, BISON, DELIS, EC-AGENTS, PACE, CREEN, IRRIIS, LiquidPub, HITIME, VIVO, GAPMINDER, GLOBALHUBS, ALLOW, ATTRACO, FRONTS, REFLECT, SOCIALNETS, SYMBRION, DREAM, SIGNAL, PEACH, EvoNet, Metaheuristics Network
- Techno-social systems, socially intelligent ICT: SOCIONICAL, CYBEREMOTION, QLECTIVES, EPIWORKS
- Collaborative Research Center SFB 555: Complex Non-Linear Processes (Berlin/Potsdam)
- Global System Dynamics (GSD)
- Global-Scale Agent-Based Models of Disease Transmission

**D6 The FuturIcT Flagship Fits the Political Agenda**

- 2020 strategy of president Barroso: give sound scientific advice to policy makers based on sound data
- New Energy Sustainable Town (NEST); CO2-neutral Masdar City (Abu Dhabi)
- Solar City (Austria), Transition Towns (UK)
- Towards a 2000 Watts society (e.g. City of Zurich)
- Ambient intelligence (from personal health care to consumer applications)
- Knowledge-driven economy
- Recent microcredit initiative of EU
- Evolution of the European political institutions
- Homeland security, relationship between crime and terrorism
- International Disaster and Risk Conference (IDRC)

    Many more initiatives could be listed here.

**D7 Milestones**

    2010: Preparation of proposal for FET Flagship pilot (Coordination Action)
    2011: Elaboration of full program and organizational issues (details of the FuturIcT Flagship)
    2012: Evaluation of the Flagship pilot: scientific program, consortium, and institutional settings
    2013: Start of Flagship
    2015: Design of new simulation and data collection concepts
    2016: First Data Collectors and Crisis Observatories in operation
    2017: First Version of Innovation Accelerator, Reputation-Based Quality Evaluation Platform
    2018: Possibilistic Computation (parallel world modelling and scenario simulation tool)
    2019: Demonstration of Reality Mining (new zero-delay sensing applications)
    2020: Demonstrators in the areas of future cities, transport, large-scale evacuation,…
    2021: Social Information Theory, Non-Equilibrium Theory of Economics
    2022: Living Earth Simulator (including Global System Dynamics Models) running

Note that each of these activities extends over several years. First demonstration systems are expected in the indicated years and are expected to trigger off subsequent large-scale research activities to enhance these systems and to make them more sophisticated.